\documentstyle[12pt]{article}

\begin{document}
%\begin{titlepage}
\title{
{\LARGE {\bf On certain cosmological relics of the early
string dynamics}}}
\author{
{\normalsize {\bf Nikolaos A. Batakis}\thanks{%
e-mail: nbatakis@cc.uoi.gr}} \\
\\{\normalsize Department of Physics, University of Ioannina} \\
{\normalsize GR-45110 Ioannina, Greece}} 
\date{}
\maketitle
\begin{abstract}
\begin{sloppypar}
\normalsize

\noindent
The tracing of cosmological relics 
from the early string dynamics may enhance the theory and provide 
new perspectives on the major cosmological problems.
This point is illustrated in a leading-order
Bianchi-type $VII_0$ string background, wherein
spatial isotropy can be claimed as such a relic, albeit a gross one.
A much finer relic, descending from 
a premordial gravitational wave, 
could be retrieved from its imprint
on the small-scale structure of the cosmic microwave background.
In spite of the absence of conventional inflation, there is no
horizon problem thanks to the presence of an equally 
fundamental mixmaster dynamics.
Implications and certain new perspectives which thus arise for the 
more general problem of cosmological mixing are briefly discussed. 

\end{sloppypar}
\end{abstract}

\addtolength{\baselineskip}{.3\baselineskip}
%\end{titlepage}
\newpage 
\section{Introduction}

Within the anticipated span of string theory, all observables 
should be viewable as relics
of the early dynamics, albeit not always easily tracable ones.
In the context of string cosmology in particular, one would be justified
to expect observables which, in addition to
measurable differences from the conventional (general relativistic)
description, would also provide 
new perspectives on the major 
cosmological problems. We do not yet have a physically acceptable
space-time realization of whatever is anticipated as a uniquely
acceptable conformal field theory or at least a reasonable
`strong-field' (namely excact to all orders in $\alpha^\prime$)
solution, which will 
illuminate the short-distance structure of space-time
\cite{1}-\cite{3}.
Nevertheless,
whatever this primordial picture may be, we know that it will
quickly (say, within the first $10^{-40}sec$) start evolving
towards the classical-gravity regime of a 4D space-time,
as exemplified by the string effective action
\cite{t},\cite{5}. Taking also into account the existing
cosmological phenomenology, one could possibly identify
string relics which would be virtually independent of the
precise short-distance formulation of the theory.
Such relics would most profitably be searched for in
the cosmic microwave background profiles and
charts, which provide {\em par excellence} 
the most accurate cosmological data \cite{cmb}.
A fundamental cosmological problem likely to be affected
by such considerations is the composite one of {\em mixing}, 
whose various facets are better known individually as 
the horizon problem, the isotropy and entropy problems,
the sensitivity to initial conditions (related to the flatness problem)
and so on \cite{4}-\cite{b}. 

The above remarks 
also summarize the motivation and general objective of this paper. 
A Bianchi-type $VII_0$ spacetime realized as a
leading-order in $\alpha^\prime$ 4D string background illustrates
the above issues, 
as it turns out to be modestly realistic and
rich in that respect. The spatial isotropy,
eventually established by
a gravity-wave driven mixmaster mechanism, can be claimed as a
gross relic of the early string dynamics.
A descendent of
that wave also survives as a much finer
relic, with valuable data for string theory and
the cosmological mixing problem.
Our main results
will be presented in sections 2,3 and further discussed in 4.

To establish notation, we recall that
the low-energy string effective
action in 4D in the Einstein frame with vanishing cosmological constant 
may be expressed as
\begin{equation}
S_{eff}=\int d^4x
\sqrt{-g}\left(R+2g^{\mu\nu}\frac{\partial_{\mu}S_+\partial_{\nu}S_-}
{(S_+-S_-)^2}\right)
\label{b},
\end{equation}
with the non-linear $SL(2,R)/U(1)$ $\sigma$-model 
variables expressed as 
$S_\pm = b\pm ie^{\phi}$
in terms of the axion/dilaton pair.
The one-loop 
beta-function equations
for conformal invariance follow from (\ref{b}) as
\begin{eqnarray}
R_{\mu\nu}+\frac{2}{(S_+-S_-)^2}\partial_\mu S_+\partial_\nu S_-&=&0, 
\label{b1}\\
\partial_\mu\left(g^{\mu\nu}\frac{\sqrt{-g}\partial_\nu S_{\mp}}
{(S_+-S_-)^2}\right)\pm
2\sqrt{-g}g^{\mu\nu}\frac{\partial_\mu S_+\partial_\nu S_-}{(S_+-S_-)^3}&=&0.
 \label{b2}
\end{eqnarray}

\section{The Bianchi-type $VII_0$ string background}

Any 4D space-time metric can always be organized as diagonal in a
properly chosen (generally non-holonomic) basis. The particular
basis of 1-forms
$\{\sigma^i,\, i=1,2,3\}$ employed here
is invariant under the left action of the
Bianchi-type $VII_0$ 3-parameter 
group of {\em transitive} Killing isometries \cite{4},\cite{5},
namely with structure constants suplied by 
the defining relations 
\begin{equation}
d\sigma^1=-\sigma^2\wedge \sigma^3,\;\;\;\;
d\sigma^2=-\sigma^3\wedge \sigma^1,\;\;\;\; d\sigma^3=0. \label{vii}
\end{equation}
The group generators are best realized 
as three independent killing vectors which define three principal
directions of anisotropy and span each one of a continuous family
of homogeneous 3D spacelike hypersurfaces $\Sigma^3$ parametrized
by $t$. More explicitly, our metric will have the form \cite{5}
\begin{equation}
ds^2=-dt^2+a_1(t)^2(\sigma^1)^2+
a_2(t)^2(\sigma^2)^2+a_3(t)^2(\sigma^3)^2,
 \label{met}
\end{equation}
wherein the metric coefficients $a_i$, just like the $b,\phi$ fields, are
functions of the proper (co-moving) time $t$ only. 
One set of explicit holonomic-coordinate $x^i$ 
realizations for the $\sigma^i$ 
resulting from the defining relations (\ref{vii})
(with $k$ an arbitrary real constant) is
\begin{equation}
\sigma^1=\;\;\cos {\frac{k}{2}x^3}dx^1+\sin {\frac{k}{2}x^3}dx^2,  \;\;
\sigma^2=-\sin {\frac{k}{2}x^3}dx^1+\cos {\frac{k}{2}x^3}dx^2,  \;\;
\sigma^3=dx^3.
 \label{viie}
\end{equation}
These expressions are subject to any transformation which preserves the
structure of (\ref{vii}). With  
the new coordinate time $\tau$ 
defined in terms of the mean radius $a$ as 
\begin{equation}
dt=a_1a_2a_3d\tau=a^{3}d\tau, \label{t}
\end{equation}
so that 
$a^{-1}da/dt$ is the mean Hubble constant of any comoving 
volume element in $\Sigma^3$, and with
a prime standing for $d/d\tau$ we may re-express 
(\ref{b1}) as
\begin{equation}
(\ln a_1^2)^{\prime\prime} +
a_1^4-a_2^4=0, \;\;\;
(\ln a_2^2)^{\prime\prime} +
a_2^4-a_1^4=0, \;\;\;
(\ln a_3^2)^{\prime\prime}- 
(a_1^2-a_2^2)^2=0. \label{VII0}
\end{equation}
The $a_i(\tau)$ are also subject to 
the initial-value equation 
\begin{equation}
(\ln a_1^2)^{\prime} 
(\ln a_2^2)^{\prime}+ 
(\ln a_2^2)^{\prime} 
(\ln a_3^2)^{\prime}+ 
(\ln a_3^2)^{\prime} 
(\ln a_1^2)^{\prime}-
(a_1^2-a_2^2)^2=A^2, \label{in}
\end{equation}
essentialy the $00$ equation in the set (\ref{b1}). 
The pair of Eqs. (\ref{b2})
can be written as
\begin{equation}
b^{\prime}+Ae^{2\phi}=0, \;\;\;\;\;\; 
\phi^{\prime\prime}+A^2e^{2\phi}=0, \label{fdp} \\ 
\end{equation}
with general solution 
\begin{equation}
b=\left(\frac{N}{A}\right)
\frac{\sinh(N\tau) +\sqrt{1-\frac{A^2}{N^2}}\cosh(N\tau)}
{\cosh(N\tau) +\sqrt{1-\frac{A^2}{N^2}}\sinh(N\tau)},\;\;
e^{-\phi}=\cosh(N\tau) +\sqrt{1-\frac{A^2}{N^2}}\sinh(N\tau), 
\label{sol}
\end{equation}
where $A,N$ are constants and where two more 
constants of integration have 
been absorbed to fix the origins of $b,\phi$. 
The totally antisymmetric tensor wherefrom $b$ descends
may be equivalently expressed as the 3-form
$H=-e^{-\phi}(db)^\ast$, namely as
\begin{equation}
H=A\sigma^1\wedge\sigma^2\wedge\sigma^3=Adx^1dx^2dx^3, \label{H}
\end{equation}
which also identifies $A$ as the magnitude of $H$ 
per unit of invariant volume in $\Sigma^3$.
We also note that
the $\sigma$-model 
contributes in (\ref{b}) as a source to the gravitational field
with the energy-momentum tensor
\begin{equation}
\kappa^2T_{\mu\nu}=
-\frac{S_+^\prime S_-^\prime}
{(S_+-S_-)^2}
\;{\rm diag}\;(1,g_{ij}).
\label{ene}
\end{equation}
The general solution
to Einstein's equations (\ref{VII0})
may be expressed as 
\begin{equation}
a_1^2=e^{2P\tau+f},\;\;\;
a_2^2=e^{2P\tau-f},\;\;\;
a_3^2=e^{2P\tau+h},
 \label{sn}
\end{equation}
still subject to (\ref{in}), with $P$ a positive constant 
and $f,h$ solutions to 
the coupled system
\begin{eqnarray}
f^{\prime\prime}+2e^{4P\tau}\sinh 2f&=&0 \label{f}, \\ 
h^{\prime\prime}-4e^{4P\tau}(\sinh f)^2&=&0. \label{h} 
\end{eqnarray}
Special solutions thereoff with $SO(2)$ and $SO(3)$ 
isotropy can be easily obtained. The latter case is realized by
the trivial solution $f=0, h=0$ of (\ref{f},\ref{h}) as
\begin{equation}
a_1=a_2=a_3=a=e^{P\tau}\sim t^{1/3}\;\;\;\;P^2=\frac{1}{12}A^2,
 \label{0}
\end{equation}
which is just a flat FRW background with axion/dilaton configurations
given by (\ref{sol}). This spatially isotropic configuration
will be of special interest to our
present considerations, not so much
as a solution by itself but rather as
the asymptotic limit of any other solution, as we will see.
Although as far as we are aware the above results are generally new, part of
them reproduces some basic aspects of the Bianchi-type $VII_0$
geometry already known in a general relativistic 
context \cite{7},\cite{b}, as well as more recent results in
string cosmology \cite{t},\cite{5}.  

\section{Mixing and relics}

The mixing process and relics
mentioned are aspects of the 
dynamics described by the solution  
(\ref{sn}) subject to (\ref{in}) together with  
the non-linear system (\ref{f},\ref{h}). This exhibits an ergodic
behavior off the initial singularity,
which has been studied as a rather typical aspect of Einstein's
equations in general, as well as in the
particular case of the Bianchi-type $VII_0$ geometry 
and its stability \cite{7}-\cite{h}. The special interest 
of (\ref{0}), associated with the trivial solution 
$f=0$ of (\ref{f}) as mentioned, lies in the fact that
it is an {\em attractor} solution, which means that it
is the only analytic one (at least in an extented domain)
and that any other solution of (\ref{f}) 
will eventually approach $f=0$ 
at $\tau\rightarrow +\infty$. 
The actual descend of any $f$
towards the $f=0$ solution
is realized following a series of initial 
adjustments to generally lower values,
whereafter $f$ enters an oscillatory evolution
(around the $f=0$
limit) with monotonically 
decreasing amplitudes and
increasing frequencies. This
behavior can be shown and stated more rigorously if we restrict
ourselves to sufficiently large times,
namely values of $t$ or $\tau$ such that 
the factor $\sinh 2f$ in
(\ref{f}) is practically equal to $2f$. Then, 
under a re-definition
of the independent variable as $x=e^{2P\tau}/P$,
(\ref{f}) is transformed to
\begin{equation}
x\frac{d^2f}{dx^2}+\frac{df}{dx}+xf=0.
\label{ff}
\end{equation}
This is the zeroth-order
Bessel equation, whose general solution is a linear combination of
the zeroth-order Bessel functions 
$J_0(x)$ and $Y_0(x)$.
From their known behavior we immediately see that their approach to zero
confirms the generally claimed pattern while,
for sufficiently large $x$ in particular, we find
\begin{equation}
f(\tau)=Fe^{-P\tau}\;\sin \left(\frac{e^{2P\tau}}{P}\right),
\label{f0}
\end{equation}
where $F$ is a constant. The behavior of $h$ can be recovred
from (\ref{h}), still subject to (\ref{in}) 
which we may re-express as
\begin{equation}
f^{\prime 2}-4Ph^{\prime}+4e^{4P\tau}(\sinh f)^2=
12P^2-A^2. \label{i2} \\ 
\end{equation}
We note, however, that we may set each side of (\ref{i2}) to zero 
if $A\ne 0$ (cf. also (\ref{0})).

The zeroes (namely the sign-changes) of $f$ give rise
to a succession of Kasner-like bounces,
in rather striking resemblance to the
mixmaster prototype \cite{6}, as
can be seen in relation to (\ref{sn},\ref{i2}). Although this
will be discussed later on, we may already
argue that the hence emerging isotropy can be claimed as a relic of
the early string dynamics by observing that 
(i) The possibility of isotropy is in the first place due to that 
dynamics, because
in a conventional general relativistic vacuum, or 
if we just turn-off the axion scalar by setting $A=0$ in
initial-value equation (\ref{in}), 
we immediately see that isotropy is forbidden.
(ii) Once possible, isotropization will eventually prevail as a
result of its association with the $f=0$ attractor solution,
seen on physical grounds as an evolution of the model
towards lower-energy-of-anisotropy configurations. 

We now observe that the metric (\ref{met}) written 
in terms of the holonomic coordinates
(\ref{viie}) together with (\ref{sn}) is 
\begin{equation}
ds^2=-dt^2+g_{ij}dx^idx^j.
\label{meth}
\end{equation}
wherein the part of the metric describing the geometry and 
evolution of the homogeneous $\Sigma^3$ sections (with $x^3=z$ 
for brevity) is 
\begin{equation}
g_{ij}=e^{2P\tau}
\left(\begin{array}{ccc}
e^f\cos^2{\frac{k}{2}z}+e^{-f}\sin^2{\frac{k}{2}z}
& (e^f-e^{-f})\cos {\frac{k}{2}z}\sin {\frac{k}{2}z}&0\\
 (e^f-e^{-f})\cos {\frac{k}{2}z}\sin {\frac{k}{2}z}&
e^f\sin^2{\frac{k}{2}z}+e^{-f}\cos^2{\frac{k}{2}z}&0\\
0&0&e^h\end{array}\right). 
\label{gij}
\end{equation}
This type of metric has been already examined in a
general-relativistic context and in particular within
a Bianchi-type $VII_0$ setting, as describing a flat  
background modulated by a standing gravitational wave or,
equivalently, as a pair of 
gravitational waves propagating in $\Sigma^3$ in 
opposite ($\pm z$) directions \cite{7}.
A similar decomposition of (\ref{gij}) will be here
directly related to the two relics mentioned
in the introduction. Indeed, in a decomposition of the form 
\begin{equation}
g_{ij}=g^{(0)}_{ij}+g^{(1)}_{ij},
\label{gspl}
\end{equation}
we may easily extract from (\ref{gij}) the flat-bacground metric 
\begin{equation}
ds_0^2=-dt^2+g^{(0)}_{ij}dx^idx^j,
 \label{meth0}
\end{equation}
with
\begin{equation}
g^{(0)}_{ij}=e^{2P\tau}
\left(\begin{array}{ccc}
\cosh f & 0&0\\
0&\cosh f & 0\\
0&0&e^h\end{array}\right), 
\label{g0}
\end{equation}
on top of which we have as a `perturbation' 
(which, of course, is neither approximate nor necessarily small)
the remaining part in (\ref{gij}), namely 
\begin{equation}
g^{(1)}_{ij}=e^{2P\tau}\sinh f
\left(\begin{array}{ccc}
\cos kz & \sin kz&0\\
\sin kz&
-\cos kz &0\\
0&0&0\end{array}\right).
\label{g1}
\end{equation}
The latter configuration is (i) traceless, 
(ii) covariantly constant and (iii) an 
eigenfunction of the wave operator,
with all operations taken with repect to 
the flat background (\ref{meth0}).
While (i),(ii) can be easily verified, to show (iii) we observe that
\begin{equation}
^{(3)}\nabla^2g^{(1)}_{ij}=
-\left(ke^{-P\tau-h/2}\right)^2g^{(1)}_{ij},
 \label{00}
\end{equation}
so that the eigenvalue in (\ref{00}) is minus the wavenumber 
($2\pi/\lambda$) squared. The amplitudes
of the wave translated within $\Sigma^3$ along $z$ are 
simultaneously differentially-rotated in the $x^1,x^2$
plane by the local $SO(2)$ element contained in (\ref{g1}).
For more direct calculations we may exploit the
invariance of (\ref{g1}) under increments of
$z$ by $\pm \pi$, expressing the resulting spacelike displacement 
by means of (\ref{meth}), to find
\begin{equation}
\lambda=\frac{2\pi}{k}e^{P\tau+h/2}=
\frac{\pi P}{k}e^{-2P\tau}l_z.
\label{wl}
\end{equation}
This result confirms our earlier identification of $-(2\pi/\lambda)^2$
as the eigenvalue 
in the rhs of (\ref{00}).
The second equality in (\ref{wl}) relates the wavelength $\lambda$ to 
$l_z$, namely to the scale of the horizon
in the $z$ direction (along which the wave is developed), defined as
\begin{equation}
l_z=a_3\int dt/a_3=
\frac{1}{2P}e^{3P\tau+h/2}=
\frac{1}{2P}a^3. \label{lz}
\end{equation}
These exact results also apply
at the $\tau\rightarrow +\infty$ limit 
where the horizon scale $l_z$ becomes of 
the order of the cosmic time $t$
as defined by (\ref{t}). More explicitly, we find 
\begin{equation}
l_z\rightarrow\frac{1}{2P}e^{3P\tau}\sim t, \;\;\;
\lambda \rightarrow \frac{2\pi}{k}e^{P\tau}
\approx \frac{2\pi}{k}a, \;\;\; 
g^{(0)}_{ij}\rightarrow e^{2P\tau}\;{\rm diag}\;(1,1,1).
\label{limits}
\end{equation}
The amplitude in (\ref{g1}) 
does not vanish (in spite of the presence of 
the $f$ factor which vanishes at that limit)
so the wave survives as the second 
(and much finer) relic mentioned 
earlier. Utilizing the asymptotic behavior of $f$
in (\ref{f0}) we find
\begin{eqnarray}
g^{(1)}_{ij}\rightarrow && Fe^{P\tau}\sin (e^{2P\tau}/P)\;
\left(\begin{array}{ccc}
\cos kz & \sin kz&0\\
\sin kz&
-\cos kz &0\\
0&0&0\end{array}\right)= \nonumber \\
&& Ft^{1/3}\sin {\frac{2\pi}{\lambda}t}\;
\left(\begin{array}{ccc}
\cos kz & \sin kz&0\\
\sin kz&
-\cos kz &0\\
0&0&0\end{array}\right),
\label{g10} 
\end{eqnarray}
which will be further discussed in the next section.

\section{Discussion and conclusions}

In addition to the ergodicity of the above solution at sufficiently
early times, we also expect higher-order
in $\alpha^\prime$ corrections as $\tau \rightarrow -\infty$.
In other words, there is an appropriately defined time $\tau_0$ near the
initial singularity, below which the present model will have
to be trancated and matched to that higher-order solution. Due
to its ergodicity in that regime, the present solution will
presumably undergo an almost dense sample of values there, virtually
including the ones to be matched at $\tau_0$ (for an at least $C^1$
junction). From that time onwards, the
evolution can be seen to 
span three epochs, delimited by the instant $\tau_{\rm iso}$ 
(to be formally
defined in the sequel)
at which isotropization has practically commenced. 

The initial epoch involves times $\tau$ 
with $\tau_0 \leq \tau \ll \tau_{\rm iso}$ 
and it is characterized by virtually 
arbitrary initial conditions,
intractable (ergodic) dynamics
which involves solutions of (\ref{f}) 
with $|f|\gg 1$, and arbitrarily large anisotropy.
The wavelength $\lambda$
is also very large compared to $l_z$, 
namely no full wavelength $\lambda$
has yet been formed within the young horizon in $\Sigma^3$
(quantified by $l_z$ and $a$, respectively, and subject to (\ref{lz})) 
as can be seen from (\ref{wl}).
According to (\ref{sn}),
the sign reversals of $f$
(later to be counted by the number of zeroes of the 
Bessel functions from (\ref{ff})) cause a
large number of Kasner-like bounces
along the principal directions of anisotropy, as mentioned earlier.
The bounces induce adequate mixing 
and isotropization while the model is been driven
towards configurations with lower energy of 
anisotropy, together with the generation of
a considerable amount of entropy, as one can deduce
from (\ref{sn}-\ref{h}) and (\ref{wl}-\ref{g10}).
This mechanism seems to offer a viable resolution not only
to the horizon problem but to the mixing problem in
general, essentially within the objectives of the
Mixmaster prototype \cite{6}-\cite{b}. 
Such a development could be a wellcome alternative, 
particularly in view of
the apparently serious difficulties faced by the more
conventional inflationary schemes \cite{12}.
 
After the initial but well before the model reaches its final epoch 
realized at $\tau\gg\tau_{\rm iso}$, there is an intermediate 
epoch chracterized by times $\tau\approx\tau_{\rm iso}$.
The wavelength $\lambda$ continues to 
grow with the overall Hubble
expansion while, simultaneously, it is shrinking with respect to 
the (faster expanding) horizon $l_z$, as seen 
from (\ref{wl}--\ref{limits}). Thus, 
the initial-epoch $\lambda\gg l_z$ relationship 
(on its way to inevitably change to $\lambda\ll l_z$ 
in in the final epoch) will pass through the equality
$\lambda=l_z$, at an instant of time which we may now formally
define as $t_{\rm iso}$, where we also 
have $|f|\approx 1$.
The emerging picture in this intermediate epoch
is that all points within the horizon 
have begun to correlate as elements of a common dynamics  
towards the formation of the
above {\em single} gravitational wave in $\Sigma^3$.
We may thus claim that this equivalence
and the associated isotropization will
have been attained {\em to first order} 
(in an approximation scheme having nothing to
do with the $\alpha^\prime$ expansion) as soon as
the first full wavelength is created within the horizon,
namely when $\lambda=l_z$ 
at $t=t_{\rm iso}$.
The second-order
approximation is realized with the formation of the second 
full wavelength within the 
horizon, and so on. 
The time required for the realization of each successive 
order is obviously decreasing, being inversely proportional to the
monotonically increasing frequency of oscillations of $f$.  

During the final epoch ($t\gg t_{\rm iso}$),
the background metric (\ref{g0}) 
evolves smoothly towards its flat isotropic limit (\ref{0}) 
and has by then completely overgrown 
the wave (\ref{g1}), now described by (\ref{g10}). 
The axion/dilaton pair, which generally develops according to (\ref{sol}), 
has driven the model to lower 
couplings \cite{8} with $e^\phi$ diminishing as
$e^{-|N|\tau}$ when $\tau\rightarrow\infty$. 
The usual cosmological parameters are given by (or
may be derived from) the results (\ref{f0},\ref{limits},\ref{g10}) and
the space-time dynamics
can be safely treated here in the adiabatic approximation. 
The amplitude, wavelength and
frequency of the wave (\ref{g10}) are easily found to be 
constant multiplied by $e^{P\tau}\sim t^{1/3}$ factors, 
carrying the overall cosmological (Hubble) expansion
common to all scales and couplings. 
The actual value of the amplitude $F$ in particular,
expected to be quite small,
is of importance because it determines the
relative magnitude of (\ref{g10}) (considered as a perturbation) 
with respect to the background metric.
Rather than being an arbitrary overall factor (if erroneously 
associated with the {\em homogeneous} equation (\ref{ff})),
$F$ actually descends from 
the non-linear (\ref{f})
and it is therefore 
expressible in terms of $A=2\sqrt{3}P$ and numerical constants.
$F$ will also specify the scale of the energy 
density stored in the relic wave,
and its relation to the 
energy of anisotropy in (\ref{ene})
wherefrom it has been effectively drawn. 

Contrasted to any anticipated {\em stochastic} 
(gravitational `black body' radiation) background
\cite{9}, 
the {\em single} gravitational wave  which survives
in the present model should have a 
distinct impact and signature in relatively
recent times, in paricular at the electromagnetic radiation
decoupling and thus on the geometry of the
hypersurface of recombination \cite{3},\cite{10}. It follows that the
parameters of such a wave could in principle be uncovered from 
multipole moments or fine modulations over the
nearly-perfect isotropic microwave background \cite{cmb},\cite{m}.
It is hoped that, in spite of its shortcommings, the present 
model is sufficiently realistic to be exploitable
in that respect. If successfully carried out, such a programme
could provide quantitative estimates on the axion/dilaton scalars
and other elements of the early string dynamics.


\newpage


\begin{thebibliography}{99}
\bibitem{1} E. Fradkin and A. Tseytlin, Nucl. Phys. B261 (1985)1;\\
C. Callan, D. Friedan, E. Martinec and M. Perry,
Nucl. Phys. B262 (1985)593.
\bibitem{2} A. A. Tseytlin, `Exact solutions of closed string theory',
preprint Imperial/TP/94-95/28, hep-th/9505052.
\bibitem{3} E.W. Kolb and M.S. 
Turner, {\em The Early Universe: Reprints}, 
 Addison-Wisley, N.Y., 1988;  {\em The Early Universe}, Addison-Wisley,
N.Y., 1990.
\bibitem{t} A. Tseytlin and C. Vaffa, Nucl. Phys. B327 (1992)443, and
additional references cited therein.
\bibitem{5} N.A. Batakis and A.A. Kehagias, 
Nuc.Phys. B 449 (1995) 248;\\
N.A. Batakis, Phys. Lett. B 353 (1995) 39.
\bibitem{cmb} G. F. Smoot et al., ApJ. Lett. 396 (1992) L1; \\
A. de Oliveira-Costa and G. F. Smoot, ApJ. 448 (1995) 447 and
additional references cited therein. 
\bibitem{4} G.F.R. Ellis and M.A.H. MacCallum, 
Commun. Math Phys. 12 (1969)108; 19 (1970)31;\\
M.A.H. MacCallum, 
`Anisotropic and Inhomogeneous Relativistic
Cosmologies', in {\em General Relativity-
An Einstein Centenary Survey}, eds.
S. W. Hawking and W. Israel, Cambridge Univ. Press, Campridge, 1979;\\
M. Ryan and L. Shepley, {\em Homogeneous Relativistic 
Cosmologies}, Princeton Univ. Press, Princeton, 1975.
\bibitem{6} C.W. Misner, Phys. Rev. Lett. 22 (1969)1071.
\bibitem{10} C.W. Misner, K.S. Thorn, J.A. Wheeler, {\em Gravitation}
Freeman, N.Y., 1973;\\
S. Weinberg, 
{\em Gravitation and cosmology}, Wiley, N.Y., 1972.
\bibitem{7} M. Demianski and L.P. Grishchuk, 
Commun. Math. Phys. 25 (1972) 2333; \\
V.N. Lukash, Sov. Phys. JETP 40 (1975) 792.
\bibitem{b} J. D. Barrow, Phys. Rep. 85 (1982)1;\\ 
J. D. Barrow and D. H. Sonoda, Phys. Rep. 139 (1986)1. 
\bibitem{h} P. Holmes, Phys. Rep. 193 (1990)137. 
\bibitem{12} R. Brustein and P.J. Steinhardt, 
Phys. Lett. B 302 (1993) 196. 
\bibitem{8} E. Kiritsis and C. Kounnas, 
`Infrared-regulated string theory
and loop corrections to coupling constants'
preprint CERN-TH/95-172, LPTENS-95/29, hep-th/9507051.
\bibitem{9} R. Brustein, M. Gasperini, M. Giovannini and G. Veneziano, 
`Relic gravitational waves from
string cosmology'
preprint CERN-TH/95-144, BGU-PH-95/06, hep-th/9507051.
\bibitem{m} E. W. Mielke and F. E. Schunk, 
Phys. Rev. D 52 (1995) 672. 

\end{thebibliography}
\end{document}